\documentclass[12pt]{article}
\usepackage{epsf,epsfig,cite}
\newcount\timecount
\newcount\hours \newcount\minutes  \newcount\temp \newcount\pmhours

\hours = \time
\divide\hours by 60
\temp = \hours
\multiply\temp by 60
\minutes = \time
\advance\minutes by -\temp
\def\hour{\the\hours}
\def\minute{\ifnum\minutes<10 0\the\minutes
            \else\the\minutes\fi}
\def\clock{
\ifnum\hours=0 12:\minute\ AM
\else\ifnum\hours<12 \hour:\minute\ AM
      \else\ifnum\hours=12 12:\minute\ PM
            \else\ifnum\hours>12
                 \pmhours=\hours
                 \advance\pmhours by -12
                 \the\pmhours:\minute\ PM
                 \fi
            \fi
      \fi
\fi
}

\def\monthname{\relax\ifcase\month 0/\or January\or February\or
   March\or April\or May\or June\or July\or August\or September\or
   October\or November\or December\else\number\month/\fi}

\def\bold#1{\setbox0=\hbox{$#1$}%
     \kern-.025em\copy0\kern-\wd0
     \kern.05em\copy0\kern-\wd0
     \kern-.025em\raise.0433em\box0 }

\def\gappeq{\mathrel{\rlap {\raise.5ex\hbox{$>$}}
{\lower.5ex\hbox{$\sim$}}}}

\def\lappeq{\mathrel{\rlap{\raise.5ex\hbox{$<$}}
{\lower.5ex\hbox{$\sim$}}}}

\def\ga{\mathrel{\raise.3ex\hbox{$>$\kern-.75em\lower1ex\hbox{$\sim$}}}}
\def\la{\mathrel{\raise.3ex\hbox{$<$\kern-.75em\lower1ex\hbox{$\sim$}}}}
\def\gev{{\rm \, Ge\kern-0.125em V}}
\def\tev{{\rm \, Te\kern-0.125em V}}
\def\beq{\begin{equation}}
\def\eeq{\end{equation}}

\def\ohsq{\Omega_{\chi} h^2}

\def\m12{m_{1\!/2}}

\newcommand{\mh}{m_{\rm h}}
\newcommand{\mH}{m_{\rm H}}
\newcommand{\mA}{m_{\rm A}}
\newcommand{\mt}{m_{\rm t}}
\newcommand{\cp}{{\cal CP}}

\begin{document}
\begin{titlepage}
\pagestyle{empty}
\baselineskip=21pt
\rightline{hep-ph/0105061}
\rightline{BNL--HET--01/14, CERN--TH/2001-116}
\rightline{DCPT/01/38, IPPP/01/19}
\rightline{UMN--TH--2004/01, TPI--MINN--01/18}
\vskip 0.05in
\begin{center}
{\large{\bf Observability of the Lightest CMSSM Higgs Boson\\
at Hadron Colliders}}
\end{center}
\begin{center}
\vskip 0.05in
{{\bf John Ellis}$^1$, 
{\bf Sven Heinemeyer}$^2$,
{\bf Keith A.~Olive}$^{1,3}$
and {\bf Georg Weiglein}$^{1,4}$}\\
\vskip 0.05in
{\it
$^1${TH Division, CERN, Geneva, Switzerland}\\
$^2${High-Energy Theory Group, Brookhaven National Laboratory, Upton,
NY~11973, USA}\\
$^3${Theoretical Physics Institute, School of Physics and Astronomy,\\
University of Minnesota, Minneapolis, MN~55455, USA}\\
$^4${Institute for Particle Physics Phenomenology, University of Durham,\\
Durham DH1~3LR, UK}\\
}
\vskip 0.5in
{\bf Abstract}
\end{center}
\baselineskip=18pt \noindent

We discuss the observability of the lightest neutral Higgs boson in the
constrained MSSM (CMSSM), with universal soft supersymmetry-breaking
parameters, at hadron colliders such as the Tevatron and the LHC. We take
account of the constraints on parameter space provided by LEP, the
measured rate of $b \to s \gamma$ decay, the cosmological relic density
$\ohsq$, and the recent measurement of $g_\mu - 2$. We normalize products
of the expected CMSSM Higgs production cross sections and decay branching
ratios $\sigma \times {\cal B}$ relative to those expected for a Standard
Model Higgs boson of the same mass. In the $h \to \gamma \gamma$ channel,
we find that $\Bigl[\sigma(gg \to h) \times {\cal B}( h \to \gamma
\gamma)\Bigr]_{\rm CMSSM} \ga 0.85 \times \Bigl[\sigma(gg \to h) \times
{\cal B}( h \to \gamma \gamma)\Bigr]_{\rm SM}$. In the $W^\pm/ {\bar t} t
+ h, h \to {\bar b}b$ channels, we find that $\Bigl[\sigma(W^\pm/ {\bar
t}t + h) \times {\cal B}( h \to {\bar b} b)\Bigr]_{\rm CMSSM} \sim 1.05
\times \Bigl[\sigma(W^\pm/ {\bar t} t + h) \times {\cal B}( h \to {\bar b}
b)\Bigl]_{\rm SM}$. We conclude that the lightest CMSSM Higgs boson should
be almost as easy to see as the Standard Model Higgs boson: in particular,
it should be discoverable with about 15~fb$^{-1}$ of luminosity at the
Tevatron or 10~fb$^{-1}$ of luminosity at the LHC.

\vfill
\vskip 0.15in
\leftline{CERN--TH/2001-116}
\leftline{May 2001}
\end{titlepage}
\baselineskip=18pt

After the completion of the LEP experimental programme, which established
a 95\% C.L. exclusion limit for the Standard Model (SM) Higgs boson of
$\mH > 113.5$~GeV and gave a hint of a possible signal of a Higgs boson
weighing 115~GeV~\cite{LEPHiggs}, the search for the Higgs boson shifts to
hadron colliders, first the Tevatron and subsequently the LHC.  There have
been many studies of Higgs production and detection, both in the SM and in
its minimal supersymmetric extension (MSSM). It just so happens that the
existence of a Higgs boson with $\mH = 115$~GeV would offer the best
prospects for the Tevatron collider and be the most challenging for the
LHC. In the SM, the conclusions have been that, if the Higgs boson $H$
weighs 115~GeV, it might be discoverable at the 5-$\sigma$ level with 15
fb$^{-1}$ of luminosity at the Tevatron collider~\cite{Tevatron} (which
might be accumulated by 2007), whilst 10 fb$^{-1}$ at the LHC
(corresponding to about one year of operation) should be enough to
discover the Higgs boson, whatever its mass up to about
1~TeV~\cite{Orsay}. In the MSSM, one expects the lightest neutral Higgs
boson $h$ to weigh $\lappeq 130$~GeV~\cite{MSSMhmass}, but the
detectability of MSSM Higgs bosons depends on other model parameters in
addition to their masses. A complete survey of MSSM parameter space would
be a very lengthy task, and attention has often focussed on particular
squark mixing scenarios~\cite{Elzbieta}. Again, the conclusions have been
encouraging for the LHC regarding the detection of at least one Higgs
boson, and there are also hopes of finding the lightest MSSM Higgs boson
$h$ at the Tevatron collider~\cite{Mrenna}. 

In this paper, we discuss Higgs observability at the Tevatron and the LHC
within the constrained MSSM (CMSSM), in which the soft
supersymmetry-breaking parameters are assumed to be universal at some high
GUT input scale~\footnote{An economical way to ensure this universality is
by gravity-mediated supersymmetry breaking in a minimal supergravity
(mSUGRA)  scenario, but there are other ways to validate the CMSSM
assumptions, including no-scale supergravity 
scenarios.}. In this case, the amount of squark mixing typically does not
coincide with that often assumed in previous analyses of MSSM Higgs
detectability at the LHC or the Tevatron~\cite{Elzbieta,Mrenna}, and the
underlying structure of
the CMSSM gives rise to a correlation between the parameters $\mA$, the
mass of the $\cp$-odd Higgs boson, and $\tan\beta$, the ratio of the
vacuum expectation values of the two Higgs doublets, which in lowest order
determine the Higgs boson phenomenology. As a new element in the
discussion of the observability of the lightest $\cp$-even Higgs boson, we
introduce the most up-to-date set of experimental and cosmological
constraints on the CMSSM parameter space, including those from
LEP~\cite{LEPHiggs}, $b \to s \gamma$~\cite{bsgexpt,bsgtheory},
cosmological dark matter~\cite{EHNOS,EFGOSi} and the anomalous magnetic
moment of the muon~\cite{BNL,ENO}. Whilst in the unconstrained MSSM the
detectability of the lightest $\cp$-even Higgs boson is not guaranteed at
the LHC even with 300 fb$^{-1}$~\cite{Elzbieta}, both the CMSSM
universality assumption and the restrictions on the CMSSM parameter space
imposed by the above constraints reduce substantially the uncertainty in
the detectability of MSSM Higgs bosons at hadron colliders, as we shall
see.

The principal mechanisms for light Higgs boson production at hadron
colliders considered in this paper are $g g \to$~Higgs~\cite{GGMN}
followed by Higgs~$\to \gamma \gamma$~\cite{EGN} and associated ${\bar t}
t$~$+$~Higgs production followed by Higgs~$\to {\bar b} b$, which are of
interest at the LHC~\cite{Orsay}, and $W^{\pm *} \to W^\pm
+$~Higgs~\cite{GNY} followed by Higgs~$\to {\bar b} b$, which is of
interest at the Fermilab Tevatron collider~\cite{Tevatron}.

{\it A priori}, the $\gamma \gamma$ signal of interest to the LHC is the
most model-dependent, since it involves loop diagrams in both the $g
g$-Higgs production vertex and the Higgs-$\gamma \gamma$ decay vertex. 
Fermion and boson loops contribute with opposite signs~\cite{EGN}, raising
the
spectre of cancellations, e.g., for particular values of the stop masses
and mixing parameters. The signal also depends inversely on the rate for
Higgs~$ \to {\bar b} b$, which is the dominant decay mode in the mass
range of interest. This can be enhanced in the MSSM, particularly for
large $\tan \beta$, offering the danger of a further suppression in ${\cal
B}(h \to \gamma \gamma)$.  On the other hand, the Higgs-${\bar t} t$
vertex is relatively model-independent, since the region of very small
$\tan\beta$ is experimentally disfavoured. Moreover, the MSSM enhancement
of the $ h {\bar b} b$ vertex actually improves the branching ratio for $h
\to {\bar b} b$, so the ${\bar t} t + h, h \to {\bar b} b$ signal at the
LHC should be relatively secure.

In the case of the $W^{\pm *} \to W^\pm +h, h \to {\bar b} b$ signature of
interest at the Fermilab Tevatron collider~\cite{Tevatron}, it is known
that the $W^\pm W^\mp h$ vertex is generically suppressed in the MSSM
relative to the SM by a factor $\sin^2 (\beta - \alpha)$.  However, as we
discuss in more detail later, this suppression does not occur in the
CMSSM, at least in the preferred parameter range that is compatible with
all the experimental and cosmological constraints. This observation,
combined with the MSSM enhancement of the $ h {\bar b} b$ vertex, suggests
{\it a priori} that the prospects for $h$ detection via this signature
should be no worse than in the SM. 

We find in this paper that, in the allowed domain of CMSSM parameter
space, $\Bigl[\sigma(gg \to h) \times {\cal B}( h \to \gamma
\gamma)\Bigr]_{\rm CMSSM}
\ga 0.85 \times \Bigl[\sigma(gg \to h) \times {\cal B}( h \to \gamma
\gamma)\Bigr]_{\rm SM}$. Values as low as 0.5 would be allowed if one
relaxed the $g_\mu - 2$ constraint, in which case $\mu <
0$ would be permitted, and furthermore abandoned the $b \rightarrow s
\gamma$ constraint, for example when $\tan \beta
= 35, A_0 = + m_{1/2}$ and $\mu < 0$. In the
$W^\pm + h, h \to {\bar b}b$ and ${\bar t} t + h, h \to {\bar b}b$
channels, we find the expected result that $\Bigl[\sigma(W^\pm/ {\bar t}t
+ h)
\times {\cal B}( h \to {\bar b} b)\Bigl]_{\rm CMSSM} \sim 1.05 \times
\Bigl[\sigma(W^\pm/ {\bar t} t + h) \times {\cal B}( h \to {\bar b}
b)\Bigr]_{\rm
SM}$, because of the enhancement in the ${\cal B}( h \to {\bar b} b)$ over
its value in the SM. 

Before describing these results in detail, we first review our treatment
of the experimental and cosmological constraints on the CMSSM parameter
space. 

There are interesting constraints from LEP on sleptons, charginos and
stops, but the most relevant is that on the Higgs boson itself. In fact,
within the CMSSM, the $b \to s \gamma$ and $g_\mu - 2$ constraints
overshadow the slepton constraint, so we do not discuss it further. The
chargino constraint is also overshadowed, except (among the cases we
study) for the choice $\tan \beta = 10, \mu > 0$. The LEP (and Tevatron
collider) constraints on stops are also important in the general MSSM
context, but not in the CMSSM discussed here. The LEP Higgs constraint
within the SM is that $\mH > 113.5$~GeV, and, as is well known, there is a
hint of a signal with mass $115.0^{+1.3}_{-0.9}$~GeV~\cite{LEPHiggs}. In
contrast to the unconstrained MSSM, for which the $Z^0 Z^0 h$ coupling is
strongly suppressed by $\sin^2 (\beta - \alpha)$ in a significant part of
the parameter space, this coupling is very close to that of the SM Higgs
for almost all possible parameter values in the CMSSM. We find a sizeable
suppression of this coupling only for $\mu < 0$, an option disfavoured by
the $g_\mu - 2$ constraint~\cite{ENO}, in small parameter regions with
large $\tan\beta$ and small $m_{1/2}$ and $m_0$. As a consequence, the SM
limit (`observed' value) can be carried over to the CMSSM for most of the
parameter
space. We allow only CMSSM parameter choices that are consistent
with $\mh > 113$~GeV in this case, so as to make some allowance for
theoretical uncertainties in the calculation of $\mh$ in the CMSSM. We
give special consideration to the range $\mh \sim 115$~GeV, but do not
impose any experimental upper limit on the CMSSM Higgs mass. For the
regions with a significant suppression of the $Z^0 Z^0 h$ coupling, we
apply the bound $\mh > 91.0$~GeV.

The theoretical uncertainties in the CMSSM Higgs mass calculations are at
present dominated by the experimental error in the mass of the top quark,
since $\delta \mh / \delta \mt = {\cal O}(1)$.  In our analysis below we
use as default $\mt = 175$~GeV, but also study the consequences if $\mt =
170$ or $180$~GeV. 


In the treatment of $b \to s \gamma$, we follow~\cite{EFGOSi} in our
implementation of NLO QCD corrections at large $\tan
\beta$~\cite{bsgtheory}. We assume the
95\% confidence-level range $2.33 \times 10^{-4} < {\cal B}(b \to s
\gamma) < 4.15 \times 10^{-4}$~\cite{bsgexpt}, and we accept all CMSSM 
parameters sets that give predictions in this range, allowing for the 
scale and model dependences of the QCD calculations. 

We assume $R$ parity conservation, so that the lightest supersymmetric
particle (LSP) is stable. The LSP is expected in the CMSSM to be the
lightest neutralino $\chi$, and may have an interesting cosmological relic
density $\ohsq$. The regions of the CMSSM parameter space allowed by
cosmology are taken from~\cite{EFGOSi}, where up-to-date results for large
$\tan \beta$ are presented. We accept CMSSM parameter sets that have $0.1
\le \ohsq \le 0.3$. Lower values of $\ohsq$ would be allowed if not all
the cosmological dark matter is composed of neutralinos. However, larger
values of $\ohsq$ are excluded by cosmology.

The final constraint that we implement is that on the possible
supersymmetric contribution $\delta a_\mu$ to the muon anomalous magnetic
dipole moment $g_\mu - 2 \equiv 2 a_\mu$, which we calculate as
in~\cite{ENO}. The signal for non-zero $\delta a_\mu = (43 \pm 16) 
\times 10^{-10}$ is considered to be a 2.6-$\sigma$ effect~\cite{BNL}, and
we
consider as preferred the 2-$\sigma$ range 
$11 \times 10^{-10} < \delta a_\mu < 75 \times 10^{-10}$. 
More conservatively, one might simply require $\delta
a_\mu \ge 0$, which is sufficient largely to exclude the $\mu < 0$
scenarios we
discuss below, that are the only ones for which we find a substantial
suppression of $\Bigl[\sigma(gg \to h) \times {\cal B}( h \to \gamma
\gamma)\Bigr]_{\rm CMSSM}$. 

For the evaluation of the cross sections and branching ratios, we use the
programs {\tt FeynHiggs}~\cite{fh}, which contains the diagrammatic
results for the complete one-loop and dominant two-loop corrections to the
Higgs-boson propagators~\cite{higgscorr}, and {\tt HDECAY}~\cite{hdecay}.
The supersymmetric parameters at the weak scale have been determined from
the parameters at the GUT scale using the two-loop renormalization-group
equations of~\cite{twoloop}. Since {\tt FeynHiggs} uses internally the
physical (i.e.\ pole) masses of the squarks, it was necessary to convert
to them from the $\overline{\rm{DR}}$ parameters (see also~\cite{chhhww}).
This was done by running all parameters in the mass matrices of the
scalar top and bottom quarks down to a renormalization scale equal to the
largest of the soft-breaking parameters in each mass matrix (when
evaluated at their own scales). The mixing matrices were then diagonalized
at this scale to yield the squark masses and mixing angles, which were
then transformed into the corresponding on-shell parameters.

We present our results in $(m_{1/2}, m_0)$ planes for the choices
$\tan \beta = 10, 50$ for $\mu > 0$, consistent with $g_\mu - 2$, and
$\tan \beta = 10, 35$ for $\mu < 0$. We do
not consider values of $\tan \beta$ below 10, since in the CMSSM the 
low-$\tan \beta$ region is severely constrained by the experimental bound
on the Higgs-boson mass~\footnote{This is in contrast to the unconstrained
MSSM, where for $\mt = 174.3$~GeV values as low as $\tan \beta = 3$ are 
allowed~\cite{lephiggsmssm}.}.

\begin{figure}
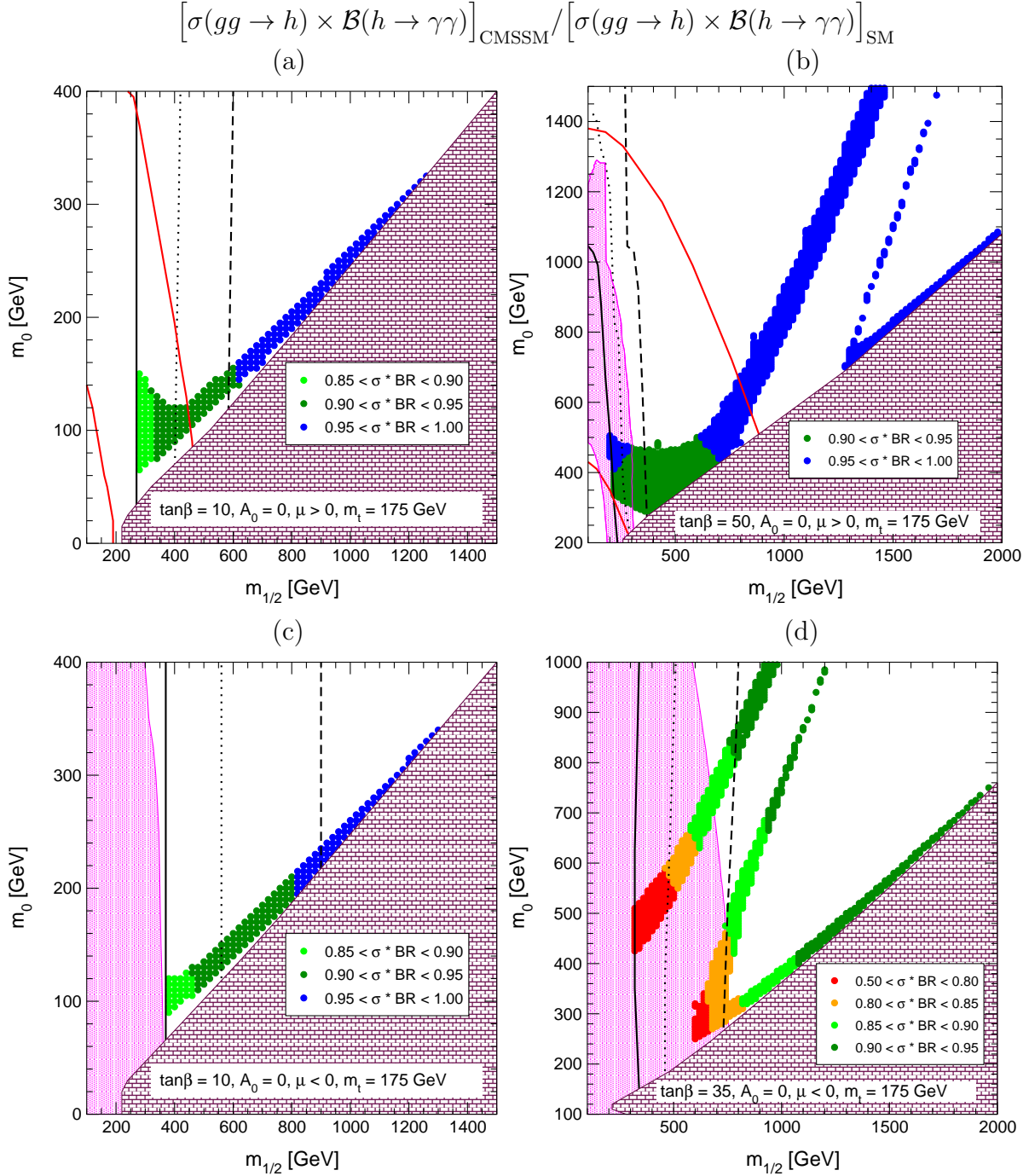


\begin{center}
$\Bigl[\sigma(gg \to h) \times {\cal B}( h \to \gamma \gamma)\Bigr]_{\rm CMSSM}
 / \Bigl[\sigma(gg \to h) \times {\cal B}( h \to \gamma \gamma)\Bigr]_{\rm
SM} $
\end{center}

\vspace{-1em}

\begin{minipage}{8in}
\hspace{4cm} (a) \hspace{7.2cm} (b) 
\end{minipage}

\vspace{.2em}

\begin{minipage}{8in}
\epsfig{file=EHOW03c.03.cl.eps,height=3.2in}
\epsfig{file=EHOW03c.09.cl.eps,height=3.2in} \hfill
\end{minipage}

\vspace{.5em}

\begin{minipage}{8in}
\hspace{4cm} (c) \hspace{7.2cm} (d) 
\end{minipage}

\vspace{.2em}

\begin{minipage}{8in}
\epsfig{file=EHOW03c.04.cl.eps,height=3.2in}
\epsfig{file=EHOW03c.08.cl.eps,height=3.2in} \hfill
\end{minipage}

\caption{\it\small The cross section for production of the lightest $\cp$-even
CMSSM Higgs boson in gluon fusion and its decay into a photon pair,
$\sigma(gg \to h) \times {\cal B}(h \to \gamma\gamma)$, normalized to
the SM value with the same Higgs mass, is given in the
$(m_{1/2}, m_0)$ planes for $\mu > 0$, $\tan\beta = 10, 50$
(upper row) and for $\mu < 0$, $\tan\beta = 10, 35$ (lower row).
In all plots $A_0 = 0$ and $\mt = 175$~GeV has been used.
The diagonal (red) solid lines in panels (a) and (b) are the $\pm 2 -
\sigma$
contours for $g_\mu - 2$: the whole parameter
space in the $\mu < 0$ plots is excluded by the $g_{\mu}-2$
constraint~\cite{BNL,ENO}.
The near-vertical solid, dotted and dashed (black) lines are the $m_h =
113, 115,
117$~GeV contours. The light shaded (pink) regions are excluded by $b
\rightarrow s \gamma$~\cite{bsgexpt,bsgtheory}. The (brown) bricked
regions are excluded since in these regions the LSP is the charged
$\tilde\tau_1$.
\label{fig:defaultgg}
}
\end{figure}

We first consider the signal for $\sigma(gg \to h) \times {\cal B}(h \to
\gamma\gamma)$, starting with the default case $A_0 = 0$ and $\mt
=175$~GeV shown in Fig.~\ref{fig:defaultgg}. Panels (a) and (b) are for
$\mu > 0$ and $\tan \beta = 10$ and $50$, respectively, where the $\pm 2 -
\sigma$ limits from $g_\mu - 2$ are shown as diagonal (red) solid lines. 
Panels (c) and (d) are for $\mu < 0$ and $\tan \beta = 10$ and $35$,
respectively, and there are no $g_\mu - 2$ contours because this sign of
$\mu$ is inconsistent with the measured value of $g_\mu - 2$. The
(near-)vertical (black) solid, dotted and dashed lines in all the panels
of Fig.~\ref{fig:defaultgg} correspond to $\mh = 113, 115, 117$~GeV, that
we take as the absolute lower limit, an indicative value and an indicative
upper limit on $\mh$, respectively~\footnote{Compared to 
earlier analyses, such as~\cite{EFGOSi}, the $m_h$
contours appear at lower $m_{1/2}$ values. This is due to an improved
Higgs mass calculation~\cite{fh,higgscorr}, where an upward shift in
$m_h$ for given $(m_{1/2}, m_0)$ of $\sim 1.5$ to $\sim 4$~GeV is
possible.}. The (pink) shaded regions are
excluded by $b \to s \gamma$, and the (brown) bricked regions are
excluded because the LSP is the lighter $\tilde \tau$ slepton. The
cosmological region where $0.1 \le \ohsq \le 0.3$ is divided into
different ranges of $\sigma(gg \to h)  \times {\cal B}(h \to
\gamma\gamma)$, relative to the SM value, that are (coloured) shaded
differently as indicated in each panel. 

We see in panel (a) of Fig.~\ref{fig:defaultgg}~\footnote{The
irregularities in the cosmological region in panel (a) etc., and the
separations between the dots in panel (b) etc. are due to the finite grid
size used in our sampling of parameter space.} that $0.85 \le
\Bigl[\sigma(gg \to h) \times {\cal B}( h \to \gamma \gamma)\Bigr]_{\rm
CMSSM}/ \Bigl[\sigma(gg \to h) \times {\cal B}( h \to \gamma \gamma)\Bigr]_{\rm
SM}  \le 1.00$ for $\mu > 0$ and $\tan \beta = 10$, once one imposes
$\mh \ge 113$~GeV and $0.1 \le \ohsq \le 0.3$. These same constraints
impose $0.90 \le \Bigl[\sigma(gg \to h) \times {\cal B}( h \to \gamma
\gamma)\Bigr]_{\rm CMSSM}  / \Bigl[\sigma(gg \to h) \times {\cal B}( h \to
\gamma \gamma)\Bigr]_{\rm SM} \le 1.00$ for $\mu > 0$ and $\tan \beta =
50$, as seen in panel (b). Here the lower limit on $m_{1/2}$ from $b \to s
\gamma$ is stronger than that from $\mh$, but does not change the lower
bound on the $h \to \gamma \gamma$ signal. Note that the BNL $g_\mu - 2$
constraint imposes $\Bigl[\sigma(gg \to h) \times {\cal B}( h \to \gamma
\gamma)\Bigr]_{\rm CMSSM}
 / \Bigl[\sigma(gg \to h) \times {\cal B}( h \to \gamma \gamma)\Bigr]_{\rm
SM}  \le 0.93 (0.96)$ for $\tan \beta = 10 (50)$.

In panel (c), for $\mu < 0$ and $\tan \beta = 10$, the $\mh$ constraint
again imposes $0.85 \le \Bigl[\sigma(gg \to h) \times {\cal B}( h \to
\gamma \gamma)\Bigr]_{\rm CMSSM}  / \Bigl[\sigma(gg \to h) \times {\cal
B}( h \to \gamma \gamma)\Bigr]_{\rm SM}  \le 1.00$, and the $b \to s
\gamma$ constraint has no impact.  On the other hand, in panel (d), for
$\mu < 0$ and $\tan \beta = 35$, we see that the $\mh$ lower limit would
allow values of $\Bigl[\sigma(gg \to h)  \times {\cal B}( h \to \gamma
\gamma)\Bigr]_{\rm CMSSM}  / \Bigl[\sigma(gg \to h) \times {\cal B}( h \to
\gamma \gamma)\Bigr]_{\rm SM}  \sim 0.50$, but this is strengthened to
$\sim 0.80$ by the $b \to s \gamma$ constraint. In this case, there is
always some suppression of the signal, and we find $\Bigl[\sigma(gg \to
h) \times {\cal B}( h \to \gamma
\gamma)\Bigr]_{\rm CMSSM}  / \Bigl[\sigma(gg \to h) \times {\cal B}( h
\to
\gamma \gamma)\Bigr]_{\rm SM} \le 0.90$ if we impose $\mH < 117$~GeV

This first survey shows (i) that the feared cancellations or other sources
of suppression in $\Bigl[\sigma(gg \to h) \times {\cal B}( h \to \gamma
\gamma)\Bigr]_{\rm CMSSM}  / \Bigl[\sigma(gg \to h) \times {\cal B}( h \to
\gamma \gamma)\Bigr]_{\rm SM} $ are relatively rare in the CMSSM, but (ii)
they may occur for $\mu < 0$ and large $\tan \beta$, in which case (iii)
they may be avoided by imposing the cosmological and experimental
constraints, notably (iv) $g_\mu - 2$ and (v) $b \to s \gamma$.

\begin{figure}

\begin{center}
$\Bigl[\sigma(gg \to h) \times {\cal B}( h \to \gamma \gamma)\Bigr]_{\rm CMSSM}
 / \Bigl[\sigma(gg \to h) \times {\cal B}( h \to \gamma \gamma)\Bigr]_{\rm
SM} $
\end{center}

\vspace{-1em}

\begin{minipage}{8in}
\hspace{4cm} (a) \hspace{7.2cm} (b) 
\end{minipage}

\vspace{.2em}

\begin{minipage}{8in}
\epsfig{file=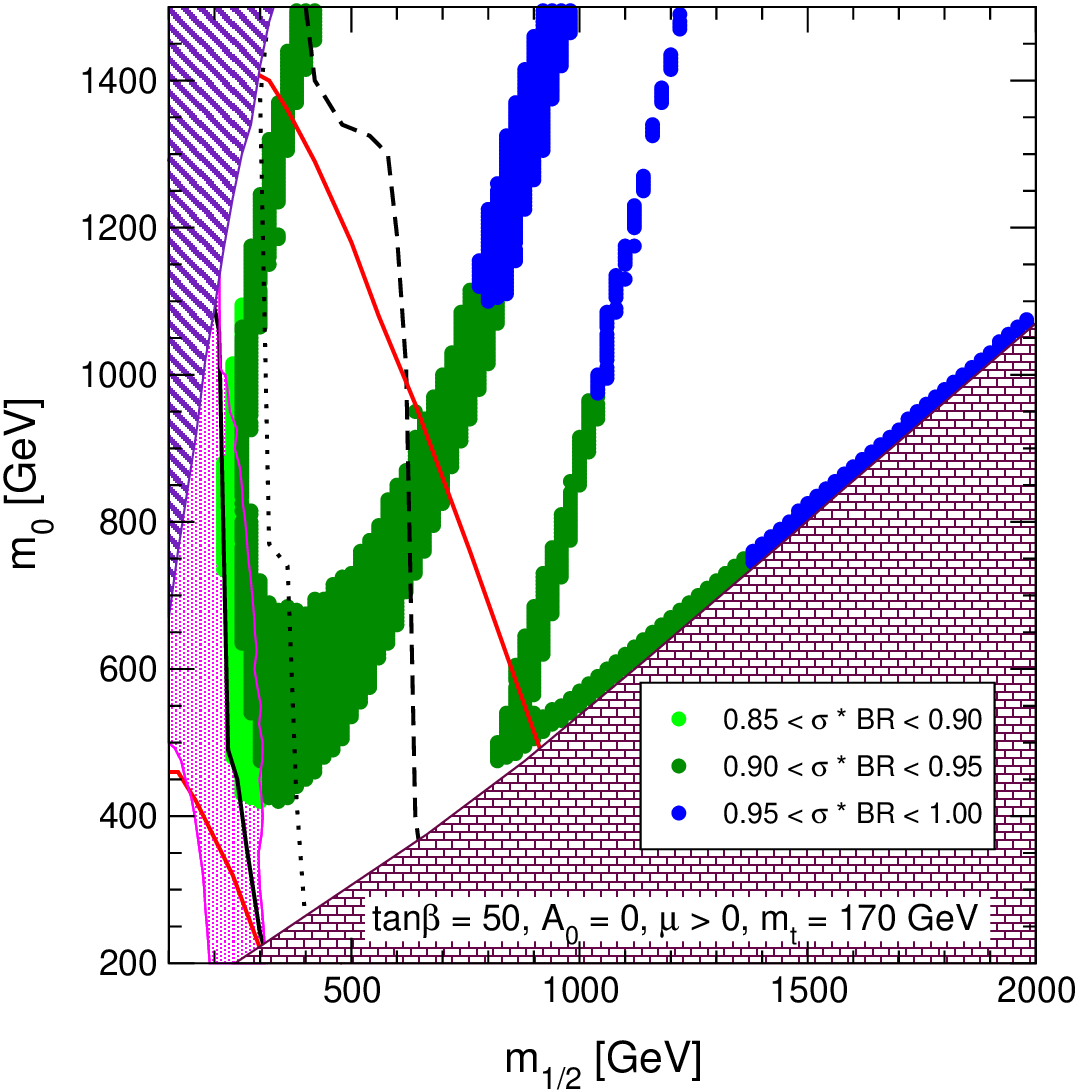,height=3.2in}
\epsfig{file=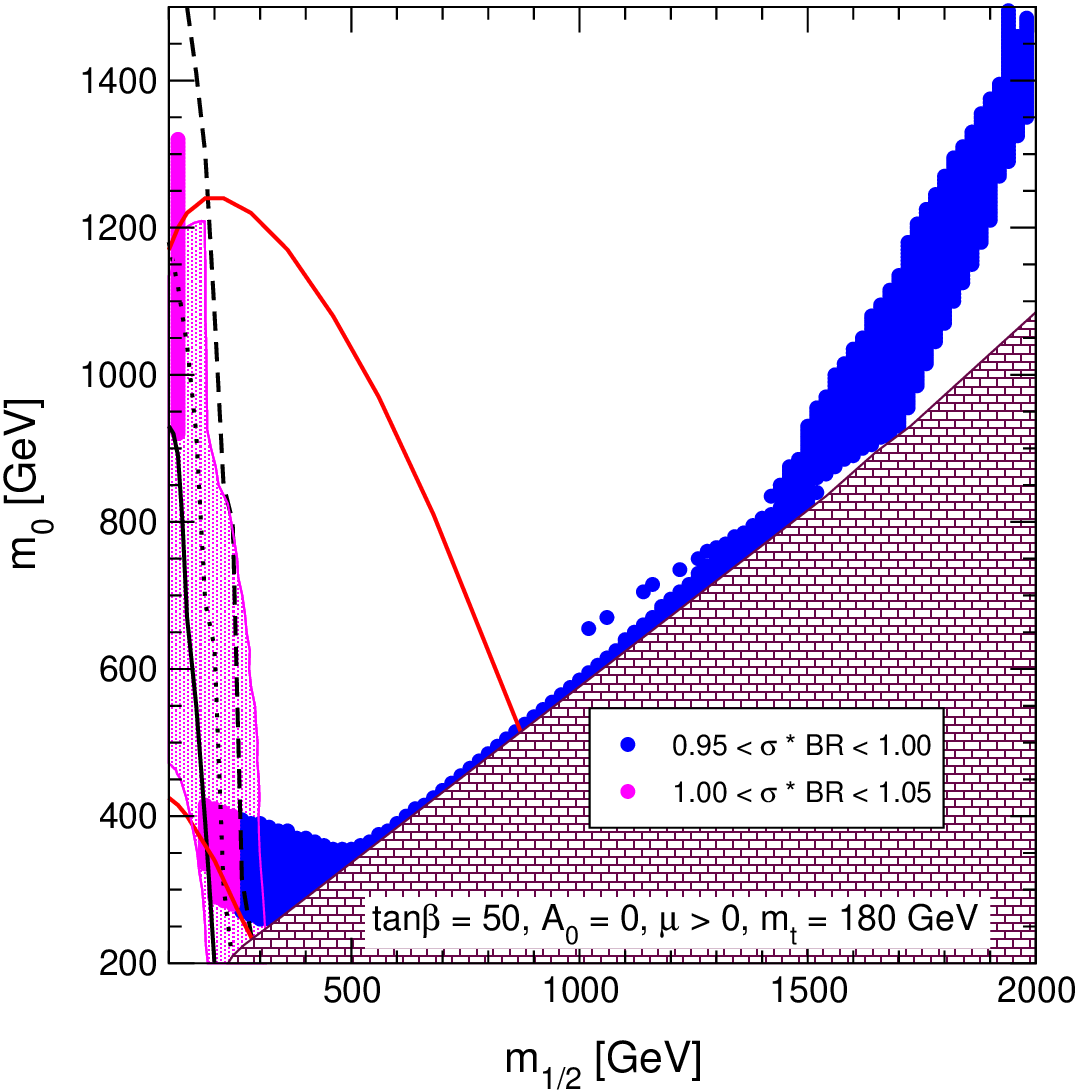,height=3.2in} \hfill
\end{minipage}

\vspace{.5em}

\begin{minipage}{8in}
\hspace{4cm} (c) \hspace{7.2cm} (d) 
\end{minipage}

\vspace{.2em}

\begin{minipage}{8in}
\epsfig{file=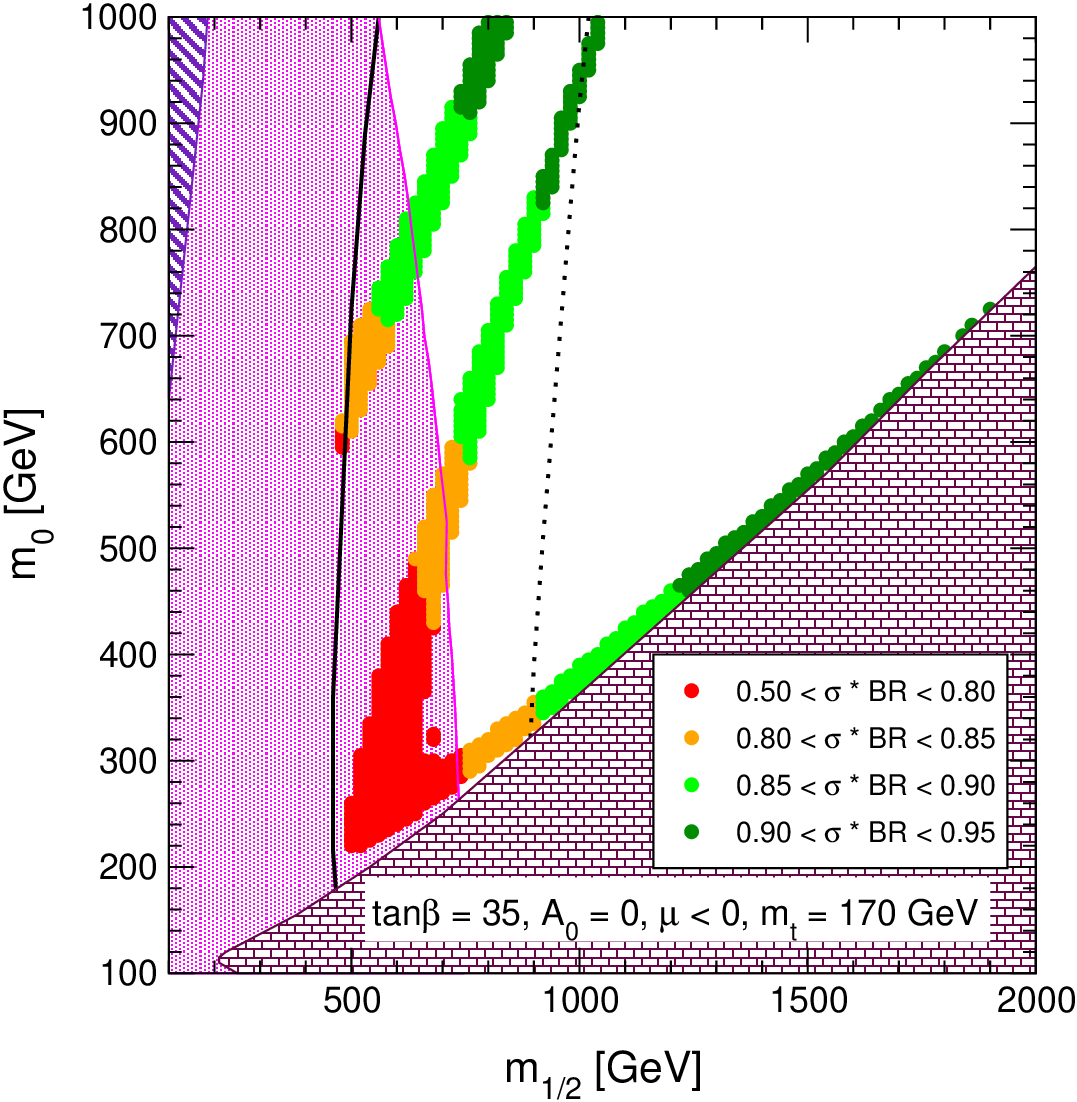,height=3.2in}
\epsfig{file=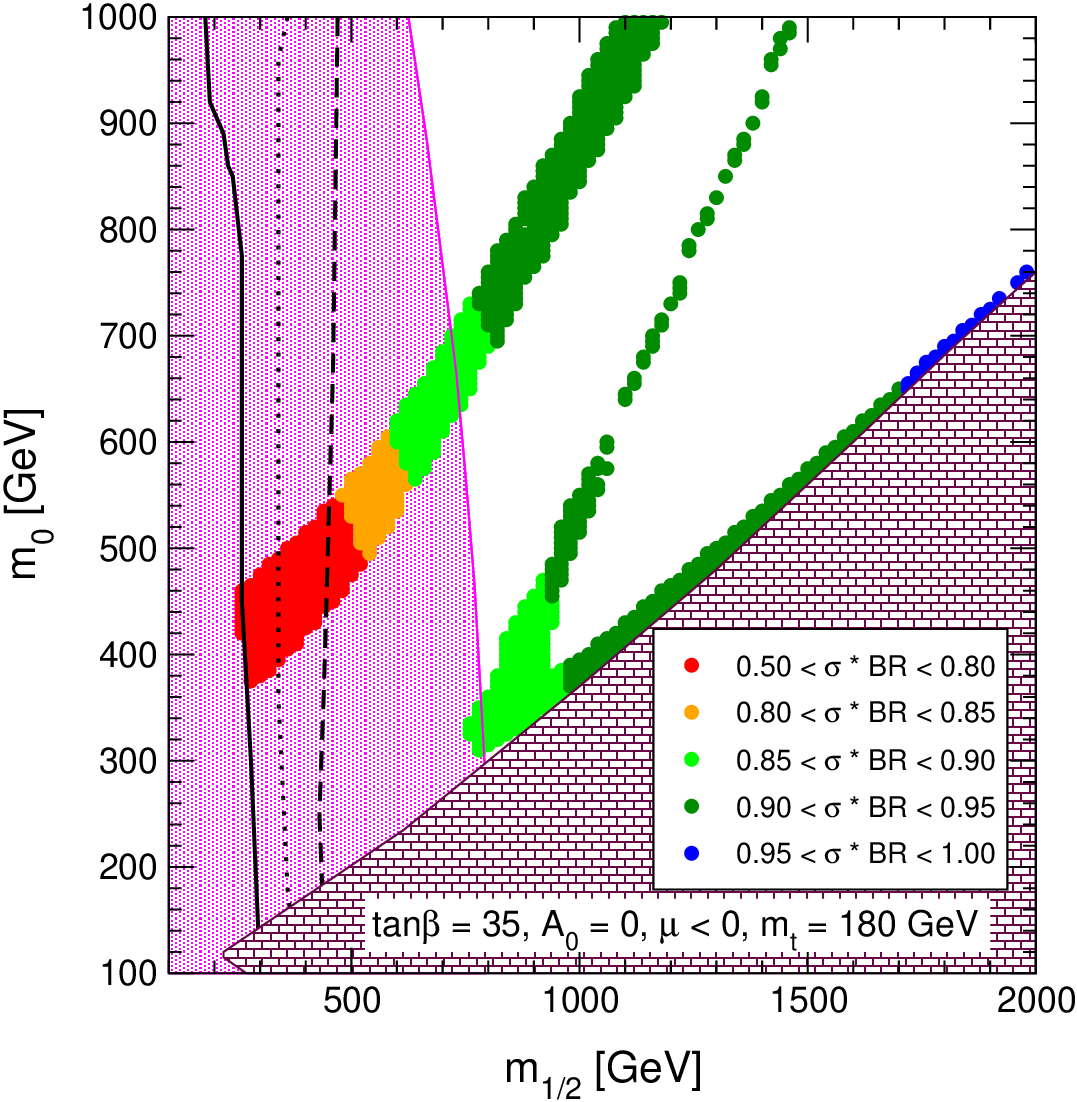,height=3.2in} \hfill
\end{minipage}

\caption{\it \small The cross section for production of the lightest $\cp$-even
MSSM Higgs boson in gluon fusion and its decay into a photon pair,
$\sigma(gg \to h) \times {\cal B}(h \to \gamma\gamma)$, normalized to
the SM value with the same Higgs mass, is given in the
$(m_{1/2}, m_0)$ planes for $\mu > 0$, $\tan\beta = 50$ and $\mt = 170,
180$~GeV (upper row) as well as for $\mu < 0$, $\tan\beta = 35$ and 
$\mt = 170, 180$~GeV (lower row).
In all plots $A_0 = 0$ has been used, and the notation is the same as in
Fig.~\ref{fig:defaultgg}. The striped regions at small values of
$m_{1/2}/m_0$ in panels (a) and (c) are excluded by the constraint of
radiative electroweak symmetry breaking.
\label{fig:varymtgg}
}
\end{figure}

We now explore the implications of varying the default parameters,
starting in Fig.~\ref{fig:varymtgg} with $\mt$. We display the cases 
$\mu > 0, \tan \beta = 50$ and $\mt = 170$~GeV in panel (a) and $\mt = 180$~GeV
in panel (b)~\footnote{The effects of varying $\mt$ for $\tan \beta = 10$ are
less important, and are not discussed here.}. We see in panel (a) that $b
\to s \gamma$ imposes $\Bigl[\sigma(gg \to h)  \times {\cal B}( h \to \gamma
\gamma)\Bigr]_{\rm CMSSM}  / \Bigl[\sigma(gg \to h) \times {\cal B}( h \to
\gamma \gamma)\Bigr]_{\rm SM}  > 0.90$, whereas the $\mh$ constraint alone would
have allowed this ratio to fall to 0.85. We see in panel (b) that this
ratio could in principle be {\it enhanced} if $\mt = 180$~GeV, although
this possibility is disallowed by $b \to s \gamma$.

In the cases $\mu < 0, \tan \beta = 35$ and (c) $\mt = 170$~GeV and (d) 
$\mt = 180$~GeV, we see again that values of $\Bigl[\sigma(gg \to h)  \times
{\cal B}( h \to \gamma \gamma)\Bigr]_{\rm CMSSM}  / \Bigl[\sigma(gg \to h)
\times {\cal B}( h \to \gamma \gamma)\Bigr]_{\rm SM} $ as low as 0.50 would be
permitted by the $\mh$ constraint, whereas the $b \to s \gamma$ constraint
strengthens this lower limit to 0.80 in panel (c) and 0.85 in panel (d) 
of Fig.~\ref{fig:varymtgg}. 

\begin{figure}
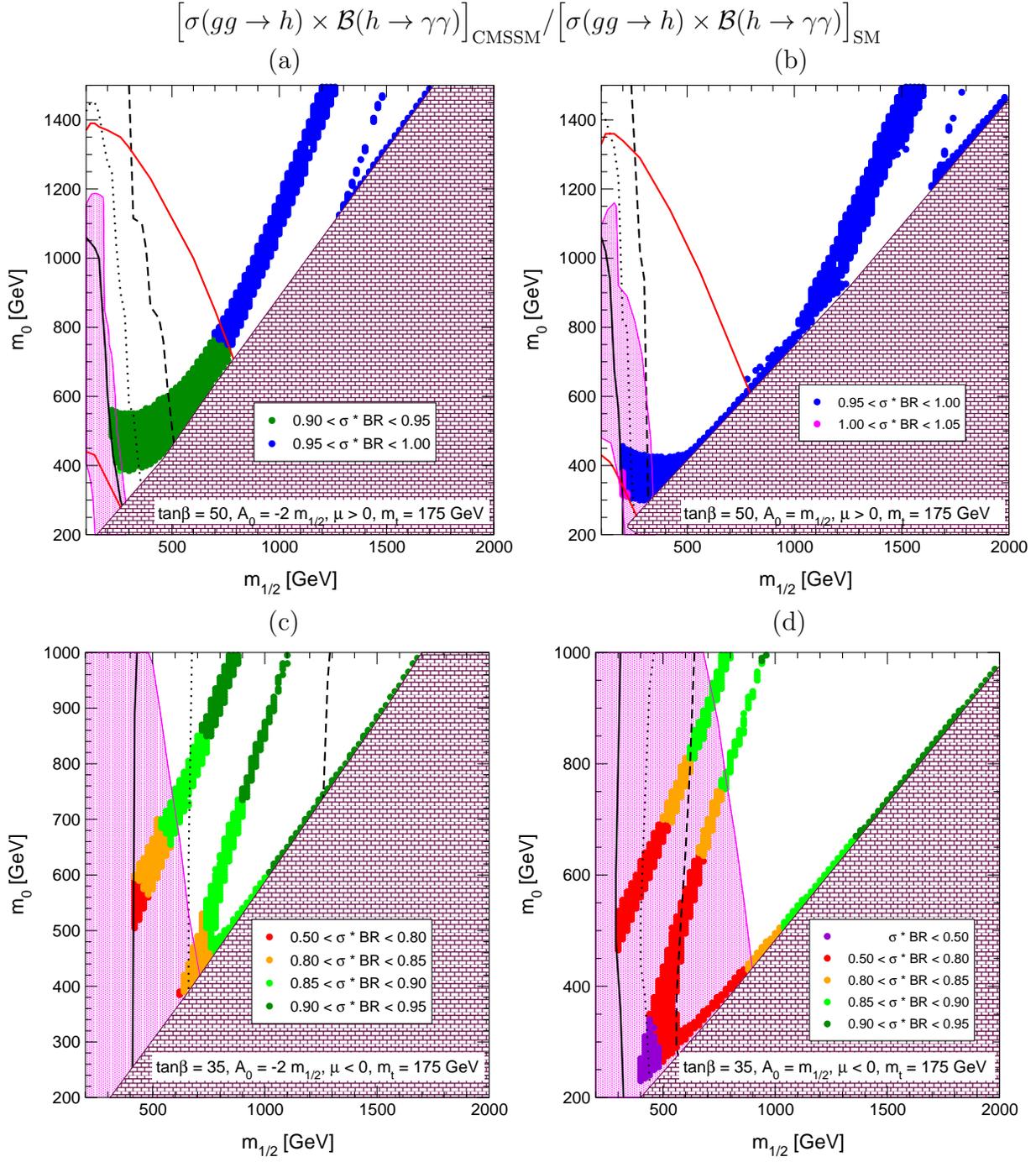


\begin{center}
$\Bigl[\sigma(gg \to h) \times {\cal B}( h \to \gamma \gamma)\Bigr]_{\rm CMSSM}
 / \Bigl[\sigma(gg \to h) \times {\cal B}( h \to \gamma \gamma)\Bigr]_{\rm
SM} $
\end{center}

\vspace{-1em}

\begin{minipage}{8in}
\hspace{4cm} (a) \hspace{7.2cm} (b) 
\end{minipage}

\vspace{.2em}

\begin{minipage}{8in}
\epsfig{file=EHOW03c.20.cl.eps,height=3.2in}
\epsfig{file=EHOW03c.26.cl.eps,height=3.2in} \hfill
\end{minipage}

\vspace{.5em}

\begin{minipage}{8in}
\hspace{4cm} (c) \hspace{7.2cm} (d) 
\end{minipage}

\vspace{.2em}

\begin{minipage}{8in}
\epsfig{file=EHOW03c.18.cl.eps,height=3.2in}
\epsfig{file=EHOW03c.24.cl.eps,height=3.2in} \hfill
\end{minipage}

\caption{\it \small The cross section for production of the lightest $\cp$-even
MSSM Higgs boson in gluon fusion and its decay into a photon pair,
$\sigma(gg \to h) \times {\cal B}(h \to \gamma\gamma)$, normalized to
the SM value with the same Higgs mass, is given in the
$(m_{1/2}, m_0)$ planes for $\mu > 0$, $\tan\beta = 50$ and $A_0 = -2
m_{1/2}, + m_{1/2}$ (upper row) as well as for $\mu < 0$, 
$\tan\beta = 35$ and $A_0 = -2 m_{1/2}, + m_{1/2}$ (lower row). 
In all plots $\mt = 175$~GeV has been used, and the notation is the same as in  
Fig.~\ref{fig:defaultgg}.
}
\label{fig:varyAgg}
\end{figure}

We now explore in Fig.~\ref{fig:varyAgg} the implications of varying
another default parameter, $A_0$, considering first the cases $\mu > 0,
\tan \beta = 50$ and (a) $A_0 = -2 \times m_{1/2}$ and 
(b) $A_0 = + m_{1/2}$~\footnote{The effects of varying $A_0$ for $\tan \beta =
10$ are also less important.}
We have also considered the case $A_0 = +2 \times m_{1/2}$, but did not
find
any significant allowed region surviving the constraints from cosmology
and 
$b \to s \gamma$. We see in panel (a)
that the $\mh$ and $b \to s \gamma$ constraints are essentially equivalent 
for this sign of $A_0$, and each impose 
$\Bigl[\sigma(gg \to h) \times {\cal B}( h \to
\gamma \gamma)\Bigr]_{\rm CMSSM}  / \Bigl[\sigma(gg \to h) \times {\cal B}( h
\to \gamma \gamma)\Bigr]_{\rm SM}  > 0.90$. In the case of $A_0 = + m_{1/2}$, 
shown in panel (b), there are again no cancellations.

The cases $\mu < 0, \tan \beta = 35$ and (c) $A_0 = -2 \times m_{1/2}$ and
(d) $A_0 = + m_{1/2}$ exhibit more variation. In the former case,
$\Bigl[\sigma(gg \to h)  \times {\cal B}( h \to \gamma \gamma)\Bigr]_{\rm CMSSM}
 /
\Bigl[\sigma(gg \to h) \times {\cal B}( h \to \gamma \gamma)\Bigr]_{\rm SM}
 <
0.80$ would be allowed by $\mh$ but not by $b \to s \gamma$. On the other 
hand, in
panel (d) for $A_0 = +m_{1/2}$, we see that values of this ratio
even below 0.50 are allowed {\it a priori}.
However, this rises to 0.80 once we impose the $b \to s
\gamma$ constraint, and we recall that this and all
$\mu < 0$ cases
are excluded by the $g_\mu - 2$ measurement. 

Apart from this possibility for reducing the $h \rightarrow \gamma
\gamma$ signal, which involves discarding some of the
principal experimental constraints on the CMSSM, we conclude that the $h
\to \gamma \gamma$ mode should be (almost) as easy to detect at the LHC as
the corresponding signal for a SM Higgs boson of the same mass. 

In view of their potential interest at the Tevatron as well as at the LHC,
we have also considered the strengths of the signals for $W^\pm/Z^0 + h, h
\to {\bar b} b$ and ${\bar t} t + h, h \to {\bar b} b$. We do not
distinguish between the
results for these two channels, as we find that, within the constraints we
impose on the CMSSM, the $W^\pm W^\mp + h$, $Z^0 Z^0 + h$ and ${\bar t} t
+ h$
couplings differ insignificantly from those in the SM. Thus the
differences in the signals from those in the SM are essentially controlled
by the differences in ${\cal B}(h \to {\bar b} b)$. It is well known that
this is generically enhanced in the MSSM relative to the SM, particularly
at large $\tan \beta$, and this is reflected in our results. 

\begin{figure}
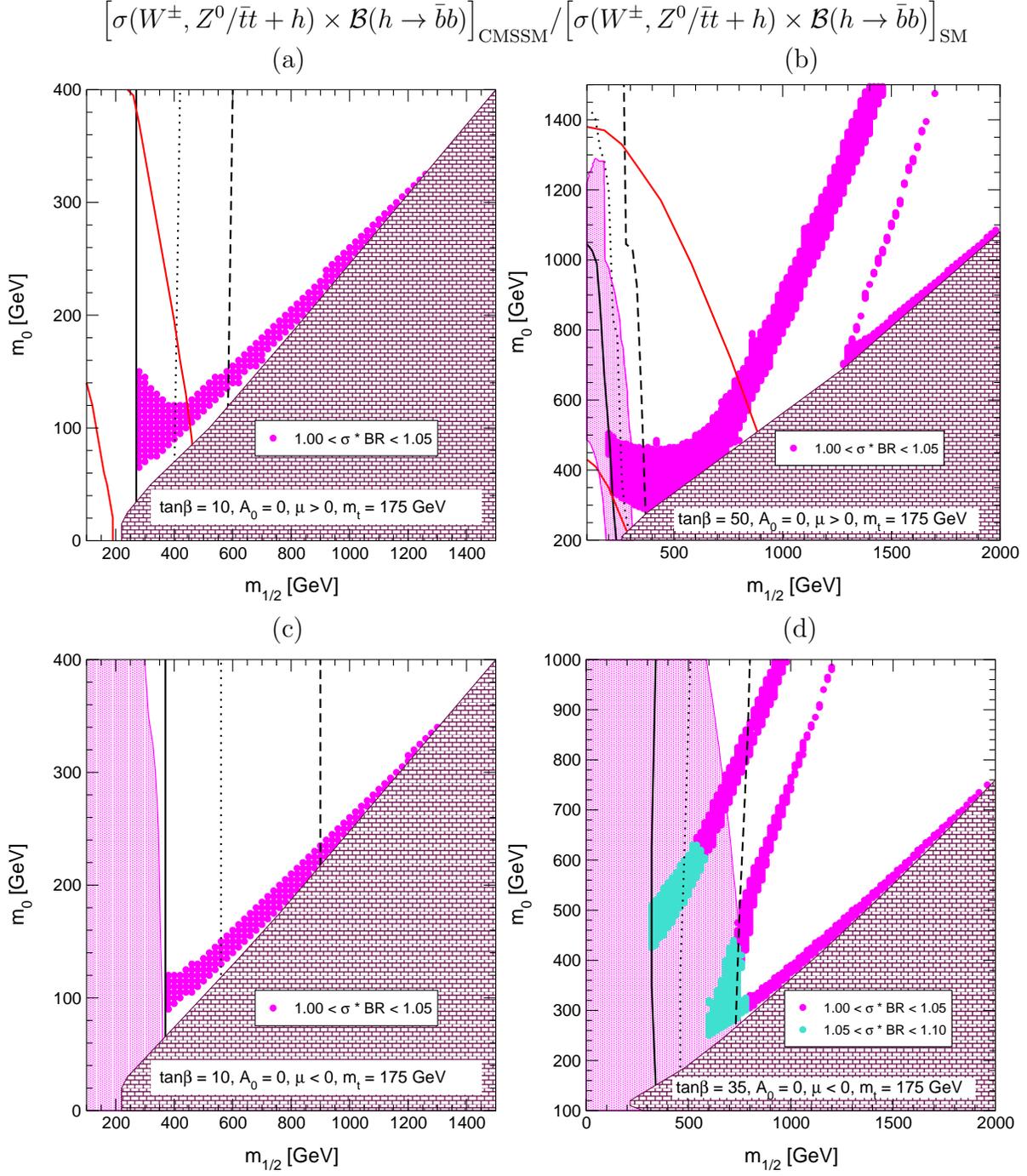


\begin{center}
$\Bigl[\sigma(W^\pm,Z^0/{\bar t} t + h) \times {\cal B}( h \to {\bar b}
b)\Bigr]_{\rm CMSSM} / 
\Bigl[\sigma(W^\pm,Z^0/{\bar t} t + h) \times {\cal B}( h \to {\bar b}
b)\Bigr]_{\rm SM}$
\end{center}

\vspace{-1em}

\begin{minipage}{8in}
\hspace{4cm} (a) \hspace{7.2cm} (b) 
\end{minipage}

\vspace{.2em}

\begin{minipage}{8in}
\epsfig{file=EHOW06c.03.cl.eps,height=3.2in}
\epsfig{file=EHOW06c.09.cl.eps,height=3.2in} \hfill
\end{minipage}

\vspace{.5em}

\begin{minipage}{8in}
\hspace{4cm} (c) \hspace{7.2cm} (d) 
\end{minipage}

\vspace{.2em}

\begin{minipage}{8in}
\epsfig{file=EHOW06c.04.cl.eps,height=3.2in}
\epsfig{file=EHOW06c.08.cl.eps,height=3.2in} \hfill
\end{minipage}

\caption{\it \small The cross section for production of the lightest $\cp$-even
MSSM Higgs boson in association with a $\bar tt$
pair or with $W^{\pm}/Z$, followed by decay into $\bar bb$, are given
normalized to the SM value with
the same Higgs mass. The results are displayed in the
$(m_{1/2}, m_0)$ planes for $\mu > 0$, $\tan\beta = 10, 50$
(upper row) and for $\mu < 0$, $\tan\beta = 10, 35$ (lower row).
In all plots $A_0 = 0$ and $\mt = 175$~GeV has been used,
and the notation is the same as in Fig.~\ref{fig:defaultgg}.
}
\label{fig:tthVVh}
\end{figure}

In Fig.~\ref{fig:tthVVh} the results for the two channels are shown for
the default case $A_0 = 0$ and $\mt = 175$~GeV. Panels (a) and (b) are for
$\mu > 0$ and $\tan \beta = 10, 50$, respectively, while panels (c) and
(d) show the case $\mu < 0$ and $\tan \beta = 10, 35$, respectively. As
expected, for all parameter values in Fig.~\ref{fig:tthVVh} we find a
slight enhancement of up to 5\% in the $W^\pm/Z^0 + h, h \to {\bar b} b$
and ${\bar t} t + h, h \to {\bar b} b$ channels compared to the SM case.
An enhancement by up to 10\% occurs for $\mu < 0$ and $\tan \beta = 35$,
but the corresponding parameter region is disfavoured by the $b \to s
\gamma$ constraint, not to mention $g_\mu - 2$.  We do not display results
for $A_0 \neq 0$ and $\mt = 170, 180$~GeV for these channels, since for
all parameter regions allowed by the $b \to s \gamma$ constraint we find
the same results as in Fig.~\ref{fig:tthVVh}, i.e.\ an enhancement
compared to the SM value of up to 5\%. 

Could one, in principle, distinguish a CMSSM Higgs boson from a SM Higgs
boson of the same mass, simply by measuring its production cross section? 
The present LHC goal for measuring luminosity at the parton-parton level
is $\pm 5$\%, and the statistical precision in the $h \rightarrow \gamma
\gamma$ channel might approach 1\%. Thus, if the theoretical error could
be neglected, there could be a 2-$\sigma$ experimental difference between
the strengths of the CMSSM and SM signals, which might be strengthened if
the luminosity precision goal could be bettered. In the case of the
$W^\pm/Z^0 + h, h \to {\bar b} b$ and ${\bar t} t + h, h \to {\bar b} b$
channels, there is a further experimental error of about 5\% associated
with the background subtractions. Thus, distinguishing between the
strengths expected in the CMSSM and the SM does not appear feasible in
these channels. 

We conclude that the lightest CMSSM Higgs boson $h$ should be almost as
easy to see as the Standard Model Higgs boson, if one accepts all the
present experimental and cosmological constraints. In particular, the
previous analyses of the prospects for Higgs searches at the LHC and
Tevatron indicate that the $h$ boson should be discoverable with about
10~fb$^{-1}$ of luminosity at the LHC~\cite{Orsay}. If its mass is about
115~GeV, i.e.\ close to the current SM exclusion bound, it is likely also
to be discoverable with 15 fb$^{-1}$ of luminosity at the Tevatron
collider~\cite{Tevatron}.

\vskip 0.5in
\vbox{
\noindent{ {\bf Acknowledgments} } \\
\noindent  
We thank 
Geri Ganis for useful information and Fabiola Gianotti for an interesting
discussion. The work of K.A.O.\ was partially
supported by DOE grant DE--FG02--94ER--40823.}

\end{document}